\documentclass[aps,prb,twocolumn,amsmath,amssymb,letterpaper]{revtex4}
\usepackage{graphicx} \usepackage{color} \usepackage{dcolumn}
\usepackage{bm}
\usepackage{hyperref}
\definecolor{violet}{rgb}{0.5,0,0.5}
\makeatletter \def\@dotsep{5} \makeatother

\newcommand{\tup}{{\hbox{$\pmb{\pmb{\boldsymbol{\blacktriangle}}}
\mkern-18mu$\raise0.15ex\hbox{\textbf{---}}}}}
\newcommand{\tdo}{{\hbox{$\pmb{\pmb{\boldsymbol{\blacktriangledown}}}
\mkern-18mu$\raise0.15ex\hbox{\textbf{---}}}}}

\begin{document}
\title{Inhomogeneous states with checkerboard order in the $t$-$J$ Model
}

\author{Chunhua Li} \author{Sen Zhou} \author{Ziqiang Wang}
\affiliation{Department of Physics, Boston College, Chestnut Hill,
MA 02467, USA}

\date{\today}
\begin{abstract}
We study inhomogeneous states in the $t$-$J$ model using an unrestricted
Gutzwiller approximation. We find that $pa\times pa$
checkerboard order, where $p$ is a doping dependent number,
emerges from Fermi surface instabilities of both the staggered flux phase and
the Fermi liquid state with realistic band
parameters. In both cases, the checkerboard order develops at wave vectors
$(\pm 2\pi/pa,0)$, $(0,\pm2\pi/pa)$ that are tied to the peaks
of the wave-vector dependent susceptibility, and is of the Lomer-Rice-Scott
type. The properties of such periodic, inhomogeneous states are discussed
in connection to the checkerboard patterns observed by STM in underdoped
cuprates.
\end{abstract}
\pacs{75.10.-b, 75.10.Jm, 75.40.Mg}
\maketitle
Spatially inhomogeneous, periodically modulated local density of states
(LDOS) has been observed recently by STM in
high-T$_c$ cuprates under a variety of conditions where
the superconductivity is weakened
\cite{Hoffman,Howald,Vershinin,Hanaguri,McElroy}.
The hope for a certain type of ordered state competing with
superconductivity has been raised in connection to the pseudogap phenomena.
The tunneling conductance map exhibits $pa\times pa$ checkerboard (CB)
patterns corresponding to wave vectors $(\pm2\pi/p,0)$, $(0,\pm2\pi/p)$,
where $p$ takes on values
between $4$ and $5$ (hereafter, we set the lattice constant $a=1$).
In and around a vortex core in
$\text{Bi}_2\text{Sr}_2\text{CaCu}_2\text{O}_{8+\delta}$ (BSCCO),
$p\sim4.3$ \cite{Hoffman}; $p\sim 4$ in optimally doped BSCCO in zero magnetic
field \cite{Howald}; $p\simeq4.7$ in underdoped BSCCO in the pseudogap phase
above $T_c$ \cite{Vershinin}; $p\simeq4$ in lightly doped oxychlorides
$\text{Ca}_{2-x}\text{Na}_x\text{CuO}_2\text{Cl}_2$ (Na-CCOC) at very
low temperatures \cite{Hanaguri}; and in substantially underdoped BSCCO,
$p\sim4.5$ were observed in the dark regions of the conductance map
\cite{McElroy}.
The nature and the origin of the LDOS modulations,
in particular the short coherence length and the role of dopants
\cite{LDC,Rice}, are unclear at present.
Several
inhomogeneous electronic states have been proposed theoretically, including
pair density waves \cite{Chen,Tesanovic};
hole Wigner crystal \cite{Fu}; a Wigner crystal of hole pairs embedded in a
d-wave resonating valence bond (RVB) state \cite{Anderson}; and
valence-bond solid with or without charge order \cite{Vojta}.
\begin{figure}[floatfix]
\includegraphics[clip,angle=90,origin=c,width=0.35\textwidth]{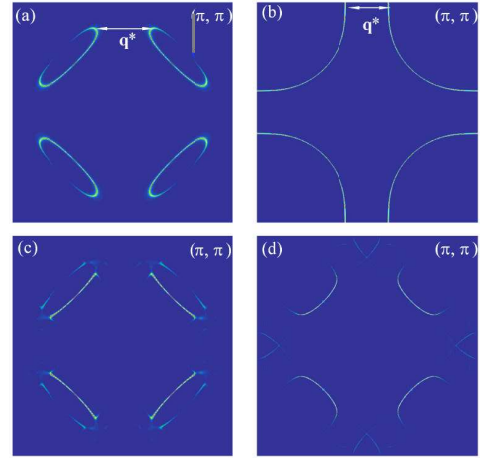}
\caption{Spectral intensity near the Fermi level in the SFP (a) and
the Fermi liquid phase with realistic band parameters (b) in the t-J model.
Those in the corresponding checkerboard ordered states are shown in (c)
and (d) where sections of the FS with high spectral
intensity connected by $\mathbf{q}^*$ are gapped.
}
\label{fig:fig1}
\end{figure}
How certain inhomogeneous electronic states with CB order
arise from the microscopic $t$-$J$ model of doped Mott insulators
is the focus of this work.
Recent density matrix renormalization group calculations
performed on small $t$-$J$ clusters found approximate CB
like patterns with strong one-dimensional stripe characters \cite{White}.
In this paper, we discuss a spatially unrestricted
Gutzwiller approximation, which allows us to study large systems in the
thermodynamic limit. We focus primarily on inhomogeneous solutions
in the nonsuperconducting phase at moderate doping where the underlying
homogeneous phase exhibits a Fermi surface (FS). Specifically, we consider
the two situations displayed in Fig.~1 for the
spectral intensity of the low energy single-particle excitations as
measured by angle-resolved photoemission (ARPES), in the staggered
flux phase (SFP) \cite{SFP} and the uniform short-range RVB state
\cite{GrilliKotliar} with realistic band parameters.
The quasiparticle scattering in the SFP (Case A in Fig.1a)
is dominated by the wave vector $\mathbf{q}^*$ connecting
the tips of the Fermi pockets with high spectral intensity.
In Fig. 1b (Case B), the sections of the large FS around
$(\pm\pi,0)$ and $(0,\pm\pi)$ are nested and scattering by the nesting
vector $\mathbf{q}^*$ is enhanced. In both cases, the wave vector dependent
susceptibility exhibits sharp peaks at $\mathbf{q}^*$ and the system
is prone to superlattice instabilities, akin to the examples discussed by
Lomer \cite{Lomer}, and Rice and Scott \cite{RiceScott} for two-dimensional
band structures with nesting and saddle-points respectively.
We show that this indeed happens in both cases, leading to inhomogeneous
CB ordered states that are lower in energy than the uniform state.
The resulting spectral intensity maps are shown in Figs. 1c and 1d
where FS sections connected by $\mathbf{q}^*$ are truncated.
In Case A, the CB order is a secondary instability of the SFP
that already exhibits a pseudogap in the LDOS.
Case B, however, is more significant. The CB order produces
a pseudogap around $(\pm\pi,0)$ and $(0,\pm\pi)$,
the antinodes of the d-wave pairing gap function, leaving behind
only ``Fermi Arcs'', in agreement with ARPES experiments \cite{FermiArc}.

In the Gutzwiller approximation, the projection of double-occupation
from the Hilbert space is partially accounted for by the statistical weight
factors, $g^t$ and $g^J$ multiplying the quantum coherent states connected
by the hopping and the exchange processes \cite{Zhang}. The latter lead to
renormalizations of the hopping and exchange parameters. The renormalized
$t$-$J$ model can be studied by decoupling the exchange term
\begin{eqnarray}
        H&=&-\sum_{i<j,\sigma}g_{ij}^t t_{ij}
\left(c_{i\sigma}^\dagger
c_{j\sigma}+h.c.\right)-\mu\sum_{i,\sigma}
c_{i\sigma}^\dagger c_{i\sigma}
\nonumber \\
        &-&\frac{3}{8}J\sum_{\langle i,j\rangle}g_{ij}^J
\left(\sum_\sigma\chi_{ij}^\ast
c_{i\sigma}^\dagger
        c_{i\sigma}+h.c.-\vert\chi_{ij}\vert^2\right)
\nonumber \\
&-&\frac{3}{8}J\sum_{\langle i,j\rangle}g_{ij}^J\left[
\Delta_{ij}^\ast\left(c_{i\uparrow}
c_{j\downarrow}-c_{i\downarrow}c_{j\uparrow}\right)+h.c.
-\vert\Delta_{ij}\vert^2\right]
\nonumber\\
&+&\sum_{i}\varepsilon_i\left
(\sum_\sigma c_{i\sigma}^\dagger c_{i\sigma}+x_i-1\right).
      \label{equation1}
  \end{eqnarray}
Here $\mu$ is the chemical potential,
$\chi_{ij}=\sum_\sigma \langle c_{i\sigma}^\dagger c_{j\sigma}\rangle$
and $\Delta_{ij}=\langle c_{i\uparrow}
c_{j\downarrow}-c_{i\downarrow}c_{j\uparrow}\rangle$ are the
the nearest neighbor bond and flux, and pairing fields,
and $t_{ij}$ is the hopping between sites $i$ and $j$.
$t_{ij}=t$ between nearest neighbors (NN); $t^\prime$ for
next NN. In the last line of Eq.~(\ref{equation1}),
$x_i=1-\sum_\sigma\langle c_{i\sigma}^\dagger c_{i\sigma}\rangle$
denotes the doped hole density at site $i$, and the averaged doping
is given by $\delta=(1/N_s)\sum_i x_i$ on a lattice of $N_s$ sites.

There are two new features in the unrestricted Gutzwiller approximation.
First, the usual Gutzwiller factors \cite{Zhang} now
depend on the local hole density,
\begin{equation}
g_{ij}^t=\sqrt{{4x_i x_j\over (1+x_i)(1+x_j)}},
\quad
g_{ij}^J={4\over (1+x_i)(1+x_j)}.
\label{gw}
\end{equation}
In essence, the local doping concentration $x_i$ is promoted to a variational
parameter \cite{LDC}. Second, as an electron hops between sites,
the renormalized bandwidth will change by ${\cal O}(1/N_s)$. However, the
kinetic energy of the occupied states changes
by an amount of order unity, and this energy difference must be
reflected in the equilibrium condition by the local fugacity
$\varepsilon_i$ in the last term of Eq.~(\ref{equation1}). The value of
$\varepsilon_i$ is determined by minimizing the energy with respect to
$x_i$,

$$\varepsilon_i=\sum_{j}\bigl[2t_{ij}{\partial g_{ij}^t\over\partial x_i}
{\rm Re}\left(\chi_{ij}\right)+
\frac{3J}{8}{\partial g_{ij}^J\over\partial x_i}
%
\left(\vert\chi_{ij}\vert^2+\vert\Delta_{ij}\vert^2\right)\bigr].
$$
In the rest of the paper,
we focus on the normal state unless otherwise noted.
\begin{figure}[htb]
\includegraphics[clip,angle=0,origin=c,width=0.48\textwidth]{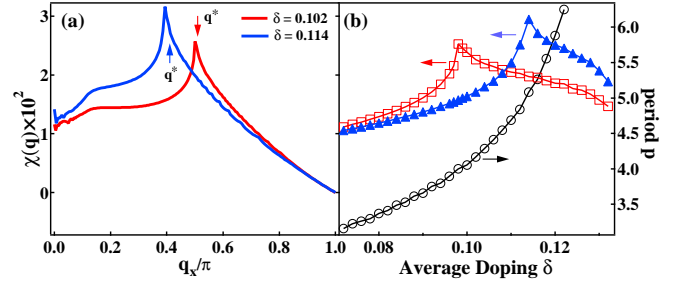}
\caption{(a) Susceptibility in the SFP in the density-bond channel along
$(q_x, 0)$. The peak positions correspond to
$\mathbf{q}^\ast = (2\pi/4, 0)$, $\mathbf{q}^\ast =(2\pi/5, 0)$
for the two doping levels.
(b) Density-bond susceptibility at fixed
$\mathbf{q}^\ast=(2\pi/4, 0)$ (squares) and
$(2\pi/5, 0)$ (trangles) as a function of doping. Also shown is
the doping dependence of inhomogeneity period $p = 2\pi/q_x^\ast$
(circles).}
\label{fig:fig2}
\end{figure}

{\it Checkerboard order in the SFP} \ We choose $t=3J$ and set $t^\prime$ 
and further neighbor hopping to zero. In this case, the uniform
normal state of Eq.~(\ref{equation1}) is the SFP for
small to moderate doping.
The spectral intensity near the Fermi level is shown
in Fig. 1a for $\delta=0.102$. To locate the wave vectors at which
Lomer-Rice-Scott instabilities may occur, we calculate the static
susceptibilities
$$
\chi_{\alpha\beta}(q)=\sum_{k\sigma,mn} \Lambda_\alpha^{mn}(k,q)
\Lambda_\beta^{nm}(k,-q)
{f(E_{k}^m)-f(E_{k+q}^n)\over E_{k+q}^n-E_{k}^m}
$$
where $\Lambda_\alpha^{mn}(k,q)$ is the coupling vertex in the density, bond,
and flux channels labeled by $\alpha$, $E_{k}^n$ is the quasiparticle band
dispersion \cite{Wangetal}. We find that all
susceptibilities exhibit peaks at $\mathbf{q}^*=(q_x^*,0), (0,q_x^*)$.
The dominant density-bond
($x$-direction) susceptibility is plotted in Fig.2a
along $(q_x,0)$. The $\mathbf{q}^*$
can be identified by the location of the sharp
peak and is doping-dependent and incommensurate in general.
If an instability occurs at $\mathbf{q}^*$, the ground
state will exhibit CB order with a periodicity $p=2\pi/{q_x^*}$,
which is shown in Fig.~2b, as a function of doping.
Note that the maximum periodicity is limited by the
stability of the SFP. For our parameters,
$p$ varies continuously between 4 and 5 in the doping range of $10-12\%$.
To search for a spatially
unrestricted Gutzwiller solution that allows static order
at $\mathbf{q}^*$ in finite
systems, it is advantageous to study doping levels at which
$\mathbf{q}^*$ takes on commensurate values.
In Fig.~2b, the susceptibility at fixed $\mathbf{q}^*=(2\pi/4, 0)$
and $(2\pi/5, 0)$ are shown as a function of doping. The
peak structures allow the choice of doping to be made.
We thus focused on an average doping $\delta=0.102$ where
$\mathbf{q}^*$ is consistent with an instability toward
$p=4$ CB order.

We next study spatially unrestricted solutions on
$80\times80$ systems with $4\times4$ and $8\times8$ supercells
wherein the $x_i$, $\varepsilon_i$, and $\chi_{ij}$ are allowed
to have spatial variations. Their values are determined by the standard
iterative solutions of the self-consistency equations.
An additional four-fold symmetry may be enforced to reduce the
number of variational degrees of freedom for faster convergence.
Remarkably, the self-consistent state with $p=4$ CB order emerges
with a significantly lower energy
compared to the uniform SFP state.
The real space CB patterns of local doping $x_i$,
staggered plaquette flux
$(-1)^i \sum_{\Box}\chi_{ij}^{\prime\prime}$,
valence bond $\chi_{i,i+\hat{x}}^\prime$, and
integrated LDOS are shown in Fig.~3 (a-d).
Their Fourier transforms show dominant peaks at
$\mathbf{q}^\ast=\left(\pm2\pi/4,0\right),
\left(0,\pm2\pi/4\right)$.
The relative modulations in the CB state and the energy
comparison to the uniform state are summarized in Table I.
Interestingly,
the charge density variations are small (less than $2\%$) while the bond and
the plaquette flux variations are more significant.
The flux and bond density wave
instabilities in the SFP of the $t$-$J$ model were
first revealed in the slave boson large-N theory \cite{Wangetal}.
These results show that weak
charge order in the SFP is a byproduct of the driving CB
order of the valence bond and the plaquette flux.
The accompanying weak charge ordering is in line with STM
observations in Na-CCOC \cite{Hanaguri}.
\begin{figure}[htb]
\includegraphics[clip,angle=0,origin=c,width=0.48\textwidth]{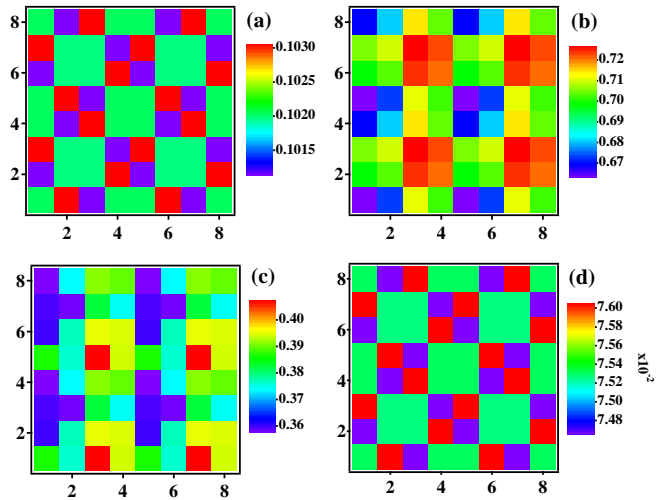}
\caption{2D maps of (a) local doping concentration, (b)
staggered plaquette flux, (c) $x$-direction bond, and
(d) integrated local density of states.}
\label{fig:fig3}
\end{figure}

It is important to note that the appearance of
the local fugacity $\varepsilon_i$ in Eq.~(\ref{equation1}) is essential for
the emergence of the CB state with lower energy than its
uniform counterpart. Indeed, without $\varepsilon_i$,
a much more inhomogeneous state can emerge as a self-consistent
solution \cite{Poilblanc},
but its energy is significantly higher than the uniform state (see Table I).
\begin{table}[htb]
\caption{\label{tab:table1} Modulation amplitudes and the energy
of the $4\times4$ and the $5\times5$ checkerboard states emerging from the
SFP and the Fermi liquid state.}
\begin{ruledtabular}
\begin{tabular}{cccccccc}
 & $x$ (\%)&bond (\%)&flux (\%)&$E_0-E_{\rm uniform} (\%)$\\
\hline
$4\times4$ & 1.87 & 12.7 & 8.76 & -1.22\\
$5\times5$ & 25.9 & 155 & $/$ &-0.167\\
Ref.\cite{Poilblanc} & 62.4 & 35.6 & 94.7 & 0.347\\
\end{tabular}
\end{ruledtabular}
\end{table}

The projected spectral function in the Gutzwiller approximation is given by
$A(i,j,\omega)=-2{\rm Im}g_{ij}^t G(i,j,\omega+i0^+)$, where $G$ is
the real-space retarded Green's function.
The Gutzwiller factor arises from projecting the matrix elements of the
electron operator between ground and excited states.
The spectral intensity at the Fermi level $A(k,\omega\simeq0)$
maps out the FS shown in Fig.~1c.
The $4\times4$ checkerboard order opens and gaps out the tips of the Fermi
pockets. The intensity is dominated by the inner branch of the remaining
segments due to the anisotropy in the SFP coherence factors.
Fine structures due to zone folding and
higher order scattering effects become visible only after the intensity scale
is reduced by four orders of magnitude.
In Fig.~4a, the band dispersion $E(k)$ extracted from the
quasiparticle peaks in $A(k,\omega)$ is shown. It retains the general
structure of the SFP, exhibiting a pseudogap in the LDOS
$\rho(r,\omega)=\sum_k e^{ikr}A(k,\omega)$ shown in Fig.~4b.
The particle-hole asymmetry is much enhanced at these parameters.
The checkerboard order induces a small energy gap at $E_F$ in the dispersion
along the $M$-$X$ direction, which is reflected in
the additional dip in the LDOS at the Fermi level.
It is worth pointing out that the suppression of the LDOS
is particle-hole asymmetric and predominately resides on the occupied side.
\begin{figure}[htb]
\includegraphics[clip,width=0.45\textwidth]{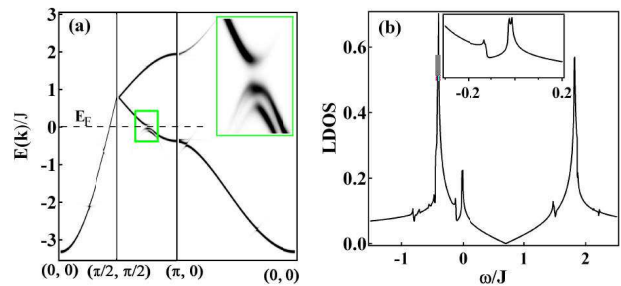}
\caption{(a) Quasiparticle dispersion in the checkerboard SFP. Inset shows
the gap near the FS. (b) LDOS site $(1,1)$. Inset shows the
LDOS near Fermi energy. }
\label{fig:fig4}
\end{figure}

{\it Checkerboard order with realistic band dispersion} \ Next we
examine whether the CB state emerges in the $t$-$J$ model
with further neighbor hopping capable of producing the dispersion
and the FS observed by ARPES \cite{NormanDing}.
The hopping integrals in Eq.~(\ref{equation1}) are chosen to
be $6.0, -2.0, 0.71, 0.7, -0.41, 0.07$ in units of $J$ from the
nearest up to the sixth neighbors respectively. The corresponding
FS in the uniform Fermi liquid was shown in Fig.~1b at doping
$\delta=0.11$. It
reveals the partially nested segments near $(\pm\pi,0)$ and
$(0,\pm\pi)$ with nesting vectors $\mathbf{q}^{\ast}\simeq(\pm 2\pi/5,0),(0,
\pm 2\pi/5)$.
The set of parameters is chosen such that the nesting properties
in the realistic band dispersion are most pronounced at low dopings.
Similar to the SFP, the calculated static, wave vector dependent
susceptibilities peak at these nesting vectors
and are most pronounced in the density-bond channel. Fig.~5a displays the
(logarithmic) divergence as $\mathbf{q}\to\mathbf{q}^*$, indicating
possible instabilities toward spatially inhomogeneous states.
On $80\times80$ systems with $10\times10$ supercells, we find that
the spatially unrestricted solution converges
to the CB state with $p=5$, consistent with the nesting vector
$\mathbf{q}^*$. The $5\times5$ pattern
is shown in Fig.~5b for the doping concentration.
The amplitudes of the variations in doping and bond (Table I)
are significantly larger than in the SFP owing to the more singular
behavior of the susceptibility. The quasiparticle dispersion obtained
from the calculated $A(k,\omega)$ near the Fermi level is shown in Fig.~5c.
The CB order opens up a sizable gap
near $(\pm\pi,0)$, $(0,\pm\pi)$ and truncates the nested segments of
the original FS as shown in Fig. 1d.

The gap near the antinodes and the
residual Fermi arcs around the nodes are
reminiscent of the pseudogap phenomenology observed by ARPES \cite{ARPES-pg}.
It is important to point out
the differences between the CB order induced pseudogap
and one that is due to a d-wave pairing gap. A finite zero-bias
conductance must remain due to
the finite ``volume'' of the residual Fermi arcs, whereas a d-wave paring gap
leads to a linearly vanishing LDOS. In Fig.~5d, the calculated
$\rho(r,\omega)$ in the $p=5$ CB state is shown.
The gap opening pushes the
van Hove peak downward, resulting in a substantial lowering of
the kinetic energy. However, the gap opens predominantly on the occupied side
leaving the spectrum particle-hole asymmetric even at low energies.
We find that this feature, also observed in the inhomogeneous SFP
in Fig.~4b, is rather generic of the CB ordered states
associated with the partial gapping of the FS. This is in contrast
to the particle-hole symmetric LDOS in a pairing-induced
pseudogap state and the conventional charge density wave state with
a fully gapped FS.
A direct comparison to the STM spectra in oxychlorides \cite{Hanaguri}
is difficult since the correlation length of the observed
CB pattern is very short ($20a$), although particle-hole
asymmetry in the tunneling spectra is apparent.
\begin{figure}[floatfix]
\includegraphics[clip,angle=0,origin=c,width=0.45\textwidth]{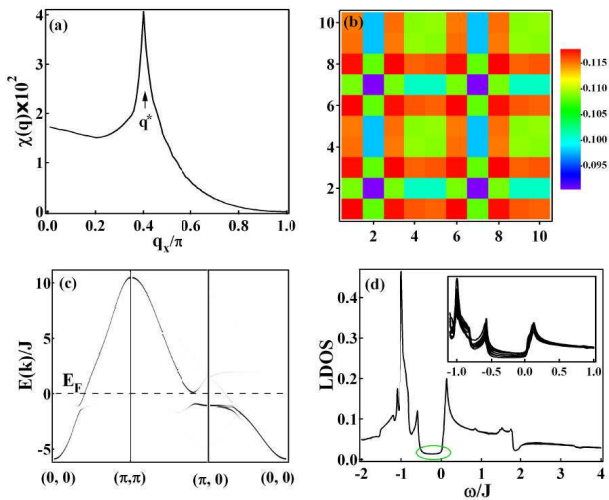}
\caption{(a) Density-bond susceptibility along $(q_x, 0)$.
(b)2D map of the local doping showing $5a\times5a$ order.
(c) Quasiparticle dispersion.
(d) Tunneling LDOS. The circle marks the gap opening just below
the Fermi level. Inset shows the spectra near Fermi level at all
different sites.}
\label{fig:fig5}
\end{figure}

In conclusion, we have shown that $pa\times pa$ ordered states emerge
in the $t$-$J$ model from both the SFP and the Fermi liquid phase
with realistic band parameters for the cuprates.
The CB states we found originate from
a generally incommensurate partial FS
instability of the Lomer-Rice-Scott variety, and are different from the
bond-ordered states \cite{Vojta} inherent of the spin-Pierls or dimerized
spin liquids at half-filling.
As such, the inhomogeneous states found in the
unrestricted Gutzwiller approximation are expected to be robust against
Gutzwiller projection of
double occupation. We have tested that the CB state is stable
in the presence of a weak pairing order parameter and
long-range Coulomb interaction. However, in the parameter regime
investigated, the uniform d-wave superconducting state
always has a lower energy. Thus, the CB
ordered state cannot coexist with d-wave pairing well inside the
superconducting phase where the FS has been gapped out except for
point zeros (nodes). Due to such competitions, the CB order discussed here
may arise in the cuprates only in the pseudogap regime or close to the
superconducting phase boundary, and may manifest in the
form of fluctuating order pinned by dopant disorder. 
Finally, the NN Coulomb interaction $V$ in the $t$-$J$-$V$ model, which 
is known to reduce the $d$-wave pairing strength while enhancing that of 
the valence bond,\cite{tjv} may lead to the coexistence of CB order 
and $d$-wave superconductivity.

We thank H. Ding, D.-H. Lee, and P.A. Lee for useful discussions.
This work is supported by DOE grant DE-FG02-99ER45747 and
ACS grant 39498-AC5M.

\end{document}